\documentclass[%
 reprint,
 amsmath,amssymb,
 aps,
]{revtex4-2}

\usepackage{graphicx}
\usepackage{dcolumn}
\usepackage{bm}
\usepackage[colorlinks=true,citecolor=blue,linkcolor=black,urlcolor=black]{hyperref}

\begin{document}

\preprint{APS/123-QED}

\title{Comment on ``Constraining the annihilating dark matter mass by the radio continuum spectral data of NGC4214 galaxy"}

\author{Volker Heesen}
\email{volker.heesen@hs.uni-hamburg.de}
\author{Marcus Br\"uggen}%
 \email{mbrueggen@hs.uni-hamburg.de}
\affiliation{University of Hamburg, Hamburger Sternwarte, Gojenbergsweg 112, D-21029 Hamburg, Germany}%

\date{Received 7 October 2020; accepted 19 November 2020}

\begin{abstract}
In their recent paper, Chan and Lee discuss an interesting possibility: radio continuum emission from a dwarf irregular galaxy may be used to constrain upper limits on the cross section of annihilating dark matter. They claim that the contributions from nonthermal and thermal emission can be estimated with such accuracy that one can place new upper limits on the annihilation cross section. We argue that the observations presented can be explained entirely with a standard spectrum and no contribution from dark matter. As a result, the estimated upper limits of Chan and Lee are at least by a factor of 100 too low.
\end{abstract}

\maketitle


Annihilation from dark matter (DM)  particles may be detected in radio continuum searches of nearby galaxies. Searches so far have concentrated on dwarf spheroidal galaxies because they have very little star formation and thus no contaminating emission at radio frequencies. So far, these searches have not provided any detection and upper limits on the radio flux density can be converted to upper limits on the annihilation cross section for various annihilation channels \citep{vollmann_20a}.

A different approach was used in \citet{chan_20a}, who analysed the radio continuum spectrum of the dwarf irregular galaxy NGC~4214. Since this galaxy, as all dwarf irregular galaxies, is star-forming, there is contamination from both thermal (free-free) and nonthermal (synchrotron) emission. In order to perform any meaningful DM detection experiment, a prior is needed on the expected radio continuum emission. The best established prior is the radio continuum-star formation (radio-SFR) relation. This relates the radio continuum luminosity to the star-formation rate in a galaxy. Using the radio-SFR relation is not an alternative method to the one used in \citet{chan_20a} but necessary even when looking at single galaxies. As the radio-SFR relation, a close corollary of the radio-far-infrared relation, is one of the most universal and tightest relations known in galaxies \citep{yun_01a}, ignoring it in order to make claims about the detection of DM is not justified.

\begin{figure}[b]
\includegraphics[width=\columnwidth]{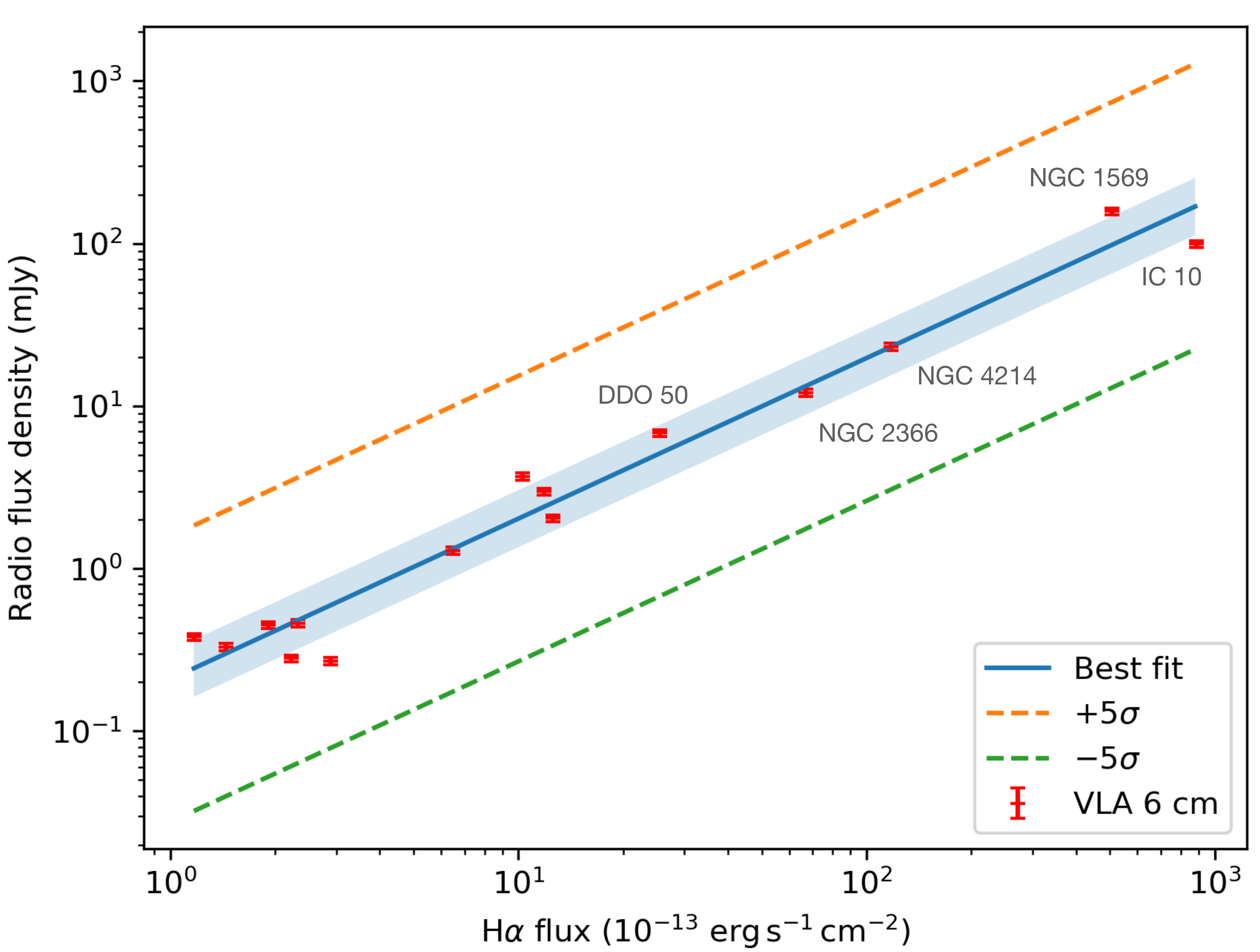}
\caption{The 6-cm radio flux density in dwarf irregular galaxies as a function of the H$\alpha$ flux. The data can be fitted by a power-law with a slope of $0.99\pm 0.05$ in accordance with a linear relation. The standard deviation is $0.18$~dex as indicated by shaded region around the best-fitting line. As can be seen, NGC 4214 is entirely consistent with the relation and no additional radio-emitting component is needed. As we argue in the text, the radio flux density would have to be $8\times$ higher than what is observed to claim a 5-$\sigma$ detection. A few brighter dwarf irregular galaxies are annotated. Data taken from \citet{hindson_18a}.}
\label{fig:radioha}
\end{figure}

In dwarf irregular galaxies, H$\alpha$ emission is a very good star-formation tracer because their dust content is low. Independent of the distance, we expect a tight correlation between the radio continuum flux density and the H$\alpha$ flux. This correlation is presented in Fig.~\ref{fig:radioha}, where NGC~4214 is consistent with the best-fitting relation. Hence, the radio continuum emission in NGC~4214 is entirely in agreement with the expected value for the given star-formation rate in this galaxy. Thus, as we would argue, this galaxy cannot provide any meaningful upper limits for the DM annihilation cross section. The main flaw in the analysis of \citet{chan_20a} is that they assume the measured flux density with its uncertainty as the prior for the expected flux density. Their assumed uncertainty is only the measured uncertainty in the radio flux density. However, such an approach is only valid if we do not expect any contribution from star formation. What really determines the uncertainty of the prior is the uncertainty in the radio-SFR relation. 


The radio continuum spectrum of NGC 4214 shows the typically concave spectral shape one expects from a constant, nonthermal radio spectral index and an increasing thermal fraction at high frequencies. Using the thermal contribution from \citet{srivastava_14a} $S_{\rm th}=20(\nu/0.1 ~\rm GHz)^{-0.1}$~mJy we can fit the data of \citet{chan_20a} (who adopted it from \citet{srivastava_14a}) adding the nonthermal contribution with $S_{\rm nt}=(120\pm 25) (\nu/\rm 0.1~GHz)^{-0.52\pm 0.08}~mJy$ with a reduced $\chi^{2}_{\nu}=1.5$. We should mention that \citet{srivastava_14a} caution against using their 150- and 325-MHz flux densities, but this does not change our conclusions. So the radio continuum spectrum can be entirely explained with a standard thermal and nonthermal spectrum without any contribution from DM. This is in fact stated in \citet{chan_20a}. Hence, it is valid to study the 6-GHz ($\lambda$6 cm) flux densities, as we have done here.

\citet{chan_20a}  fix the nonthermal and thermal contributions and add a possible DM signal in order to derive new upper limits for the annihilation cross section. They assume a constant spectral index for the cosmic-ray synchrotron spectrum and deviations from this assumed spectrum are attributed to a DM signal. However, any deviation from a power-law spectrum can easily be explained by processes such as cosmic-ray injection, cosmic-ray transport and energy losses. These uncertainties are ignored when making a fit thus invalidating claims of a DM signal by \citet{chan_20a}.
The flaw in their analysis is that they use the error of the integrated flux densities as the error for their 1$\sigma$ DM detection signal. In principle, the flux densities have a typical error of 10 percent. Following the argument in \citet{chan_20a}, adding 50 percent to the flux density, as they do, would then amount to a 5$\sigma$ detection of a DM annihilation signal. This is shown in their Fig.~3. This is of course not correct. The significance of the DM signal does not in this case primarily depend on how accurately the flux densities are measured. The dominant uncertainty lies in the expected radio luminosity for a given star-formation rate. Hence, what really determines the significance of the DM detection is the deviation from the radio-SFR relation. More specifically, the uncertainty in the nonthermal synchrotron spectrum is the main uncertainty, since for the thermal part H$\alpha$ emission is an excellent proxy.

\citet{hindson_18a} showed that the uncertainty for the radio-SFR relation in dwarf irregular galaxies is only $0.2$~dex, equivalent to $\pm$50 percent. Nevertheless, that means that a 5$\sigma$ detection requires a deviation of 1~dex, equivalent to a factor of 10, from the radio-SFR relation. A similar excess is needed for the flux densities as we show in Fig.~\ref{fig:radioha}. Only if such an excess is found, one can start to speculate about additional emission mechanisms such as a signal from DM annihilation. In short, rather than changing the flux density of NGC~4214 by 50 percent ($\approx$50 mJy) for a 5$\sigma$ detection, the flux density needs to increase by a factor of 10 ($\approx$900~mJy), a factor of nearly 20 larger. The upper limits for the annihilation cross section would increase by the same factor. Of course, one would then still need to look for other sources of contamination such as from background radio galaxies, but we leave aside this complication for now.

Another shortcoming of the analysis in star-forming dwarf irregular galaxies is that the cosmic-ray residence time may be  limited due to outflows and winds. For NGC~4214, \citet{kepley_11a} estimate  a residence time of 10~Myr as indicated by the flat radio spectral index. The corresponding diffusion coefficient $D=\rm (1~kpc)^2/(10~Myr)$ is with $D=3\times 10^{28}~\rm cm^2\,s^{-1}$ a factor of $\sim$100 larger than what \citet{chan_20a} used in their analysis. Even using a lower value of $10^{27}~\rm cm^2\,s^{-1}$ would suppress the radio continuum luminosity by another factor of 10 since the cosmic-ray energy density is suppressed. Taken together, we estimate that the derived upper limits of \citet{chan_20a} for the annihilation cross section are by at least a factor of 100 too low. 

In summary, the complex relations that regulate the radio continuum emission in galaxies with star formation mean that they are not suitable for a DM search in the radio continuum. This is particularly the case for dwarf irregular galaxies that cannot hold on to their cosmic-ray electrons. The resulting suppression of the synchrotron emission leads to an increased scatter of the nonthermal radio-SFR relation \citep{hindson_18a}. Better targets for a DM search are dwarf spheroidal galaxies, which lack this complication.

\begin{acknowledgments}
 This work is supported by the Deutsche Forschungsgemeinschaft (DFG, German Research Foundation) under Germany’s Excellence Strategy – EXC 2121 Quantum Universe – 390833306.
\end{acknowledgments}

\bibliography{apssamp}

\end{document}